\documentclass{PoS} 
 
\usepackage[dvips]{graphics,psfrag}
\usepackage{epsfig}
\usepackage{psfig}
\usepackage{colordvi}
\usepackage{exscale}

\PoS{PoS(LAT2005)148} 

  \newcommand{\bq}{\begin{equation}}
  \newcommand{\eq}{\end{equation}}
  \newcommand{\bqa}{\begin{eqnarray}}
  \newcommand{\eqa}{\end{eqnarray}}

  \newcommand{\bc}{\begin{center}}
  \newcommand{\ec}{\end{center}}
  \newcommand{\bmp}{\begin{minipage}}
  \newcommand{\emp}{\end{minipage}}

  \newcommand{\bit}{\begin{itemize}}
  \newcommand{\eit}{\end{itemize}}

\def\lsim{\raise0.3ex\hbox{$<$\kern-0.75em\raise-1.1ex\hbox{$\sim$}}}
\def\gsim{\raise0.3ex\hbox{$>$\kern-0.75em\raise-1.1ex\hbox{$\sim$}}}

\def\PBP{{\langle \bar\psi\psi \rangle }}
\def\ns{N_{\sigma}}

\title{The chiral transition of $N_f=2$ QCD with fundamental
       and adjoint fermions} 
 
\ShortTitle{The chiral transition for fundamental and adjoint fermions} 
 
\author{\speaker{J.~Engels}\thanks{We thank Doug Toussaint for his help 
with the MILC code.}~, S.~Holtmann, and T.~Schulze\\ 
        Fakult\"at f\"ur Physik, Universit\"at Bielefeld,
        D-33615 Bielefeld, Germany\\ 
        E-mail: \email{engels@physik.uni-bielefeld.de},
                \email{sven.holtmann@depfa.com},
                \email{tschulze1973@gmx.de} } 
 
\abstract{
We study QCD with two staggered Dirac fermions both in the fundamental
($QCD$) and the adjoint representation ($aQCD$) near the chiral
transition. The aim is to find the universality class of the chiral
transition and to verify Goldstone effects below the transition. We
investigate $aQCD$, because in that theory the deconfinement and the
chiral transitions occur at different temperatures $T_d<T_c$.
Here, we show that the scaling behaviour of the chiral condensate 
in the vicinity of $\beta_c$ is in full agreeement with that of the
$3d$ $O(2)$ universality class. In the region $T_d<T<T_c$ we confirm the
quark mass dependence of the chiral condensate which is expected due to
the existence of Goldstone modes like in $3d$ $O(N)$ spin models. For
fundamental $QCD$ we use the $p4$-action. Here, we find 
Goldstone effects below $T_c$ like in $aQCD$ and the $3d$ $O(N)$ spin 
models, however no $O(2)/O(4)$ scaling near the chiral transition 
point. The result for $QCD$ may be a consequence of the coincidence
of the deconfinement transition with the chiral transition. } 
 
\FullConference{XXIIIrd International Symposium on Lattice Field Theory\\ 
                 25-30 July 2005\\ 
                 Trinity College, Dublin, Ireland} 

\begin{document} 
 
\section{Introduction}
 
We study $QCD$ with two staggered Dirac fermions both in the fundamental
($QCD$) and the adjoint representation ($aQCD$) near the chiral
transition. The intention is to find the universality class of the
transition and to verify Goldstone effects below the transition. For
ordinary $QCD$ with staggered fermions the prediction of the
three-dimensional $O(2)$ universality class (the $O(4)$ class in the 
continuum theory) could up to now not be confirmed.
We investigate in addition $aQCD$, because in that theory the deconfinement
and the chiral transitions occur at different temperatures with $T_d<T_c$
(that is $\beta_d<\beta_c$) \cite{Karsch:1998qj}. The chiral transition can
therefore be studied without interference. The comparison of $QCD$ data 
with the critical behaviour of $O(N)$ spin models requires the following 
identifications of $QCD$ variables to $O(N)$ variables: the chiral
condensate corresponds to the magnetization $M$ and the quark mass
$(m_qa)$ to the magnetic field $H$. Instead of the temperature we use
$\beta=6/g^2$.
 
\section{$QCD$ with adjoint fermions ($aQCD$)}
 
The action of $aQCD$ which we use is \cite{Karsch:1998qj}
\bq 
S=\beta S_G(U^{(3)}) + \sum_{x,y} \bar\psi_x M(U^{(8)})_{x,y}\,
\psi_y~. 
\eq
Here, the gluon part $S_G(U^{(3)})$ is the usual Wilson one-plaquette
action, but the fermions are in the 8-dimensional adjoint representation
of color $SU(3)$. The standard staggered fermion matrix $M$ depends
correspondingly on $U^{(8)}$ instead of $U^{(3)}$. The links $U^{(8)}$ are
{\em real} because
\bq 
U^{(8)}_{ab} ={1 \over 2} {\rm tr}_3\left[ \lambda_a U^{(3)}
\lambda_b U^{(3)\dagger}\right]~. 
\eq
The fermion action does not break $Z(3)$ center symmetry and the Polyakov 
loop $L_3$ is therefore order parameter for the deconfinement transition.
In the continuum $aQCD$ contains an $SU(2N_f)$ chiral symmetry which breaks
to $SO(2N_f)$ for $T<T_c$. This is because here $N_f$ Dirac fermions 
correspond to $2N_f$ Majorana fermions. For $N_f=2$ we have $SU(4)$-symmetry
which breaks to $SO(4)$. The corresponding continuum transition has been
studied in Ref.\ \cite{Basile:2004wa} with renormalization-group methods.
On the lattice an $O(2)$-symmetry remains for
staggered fermions.

Our simulations \cite{Engels:2005te} were done on $\ns\times 4$ lattices
with $\ns=8$, 12
and 16 and a fixed length $\tau=0.25$ of the trajectories. We used 900-2000
trajectories for measuring the chiral condensate $\PBP$, the Polyakov loop
$L_3$ and the disconnected part $\chi_{dis}$ of the susceptibility.  
 
In Ref.\ \cite{Karsch:1998qj} the deconfinement transition point was found
at $\beta_d=5.236(3)$ - the same value as in $QCD$(!), but here it is of
first order. The usual strategy to locate the chiral transition
point $\beta_c$ is to extrapolate the line of peak positions $\beta_{pc}$
of the susceptibility to $m_qa=0$ with
\bq 
\beta_{pc} = \beta_c +c (m_qa)^{1/\Delta}~, 
\eq
where $\Delta=\beta_m \delta$ is a product of critical exponents. On the
$8^3\times 4$ lattice we find for $m_qa=0.005,0.01,0.02$ the values
$\beta_{pc}=5.73(8),5.74(6),5.75(10)$. With $\Delta$ from $O(2)$
we obtain $\beta_c\!=\!5.7(2)$. A better method \cite{Engels:2005te}
is to expand the scaling ansatz $\PBP=m^{1/\delta}f(z)$ at at small $|z|$.
Here, $m=m_qa/(m_qa)_0$, $\beta_r=(\beta-\beta_c)/\beta_0$ are the reduced
field and temperature and $z=\beta_r m^{-1/\Delta}$. The result is
\bq 
\PBP = (m_q a)^{1/\delta}\left\{ d_c +  d_c^1 (\beta-\beta_c) 
(m_q a)^{-1/\Delta} + \dots \right\}~. 
\label{fitbc}
\eq
Fits to only the first term at fixed $\beta$ are best for $\beta\in [5.6,5.7]$. 
If $O(2)$ exponents and two terms are used one finds a unique zero of 
the parameter $ d_c^1 (\beta-\beta_c)$ as a function of $\beta$ at the 
rather precise value $\beta_c = 5.624(2)$. 

\begin{figure}[t]
\begin{center}
      \epsfig{bbllx=63,bblly=265,bburx=516,bbury=588,
       file=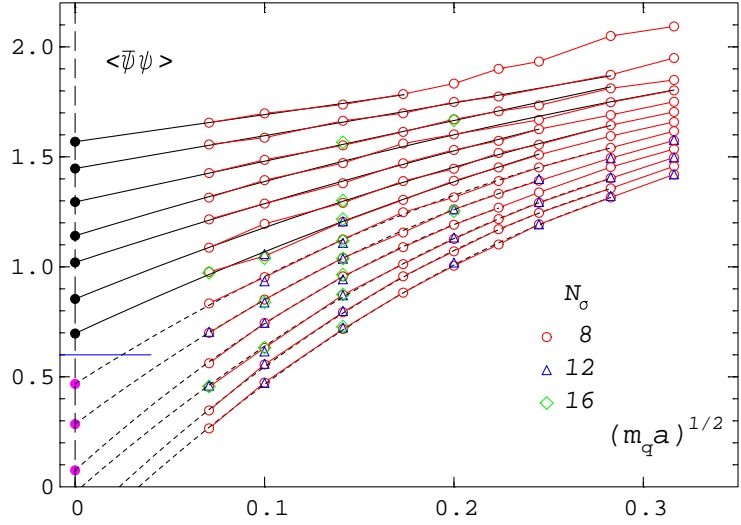,width=90mm,angle=0} 
\end{center}
\vspace*{0.2cm}
\caption{The chiral condensate $\PBP$ of $aQCD$ as a function of 
$(m_qa)^{1/2}$ for all $\beta$-values between 5.3 (highest values) and
5.9 (lowest values) from  $\ns^3 \times 4$ lattices with $\ns=8$ (circles),
12 (triangles) and 16 (diamonds). The black lines are fits with ansatz
(2.5), the filled circles denote the extrapolations to $m_qa =0$.}
\label{fig:golda}
\end{figure}

Due to the existence of massless Goldstone modes for all
$T<T_c$ the susceptibility $\chi_L$ of $3d$ $O(N)$ spin models diverges for 
$H\rightarrow 0$ as $\chi_L \sim H^{-1/2}$. If $aQCD$ behaves effectively
as such a model, a similar divergence is expected. In terms of the
chiral condensate we must have then
\bq 
\PBP(\beta,m_qa) = \PBP(\beta,0) +c_1(\beta) (m_qa)^{1/2}
 +c_2(\beta) (m_qa)+ \dots 
\label{pbpgold}
\eq
Corresponding fits are shown in Fig.\ \ref{fig:golda} for $\beta=5.3$ 
to $\beta=5.9$ with $\Delta\beta=0.05\,$.  The blue line separates fits 
above and below $\beta_c$. We see that the model expectations are met well
by the data.

\begin{figure}[t]
\setlength{\unitlength}{1cm}
\begin{picture}(13,7)
\put(0.5,0.5){
   \epsfig{bbllx=127,bblly=264,bburx=451,bbury=587,
       file=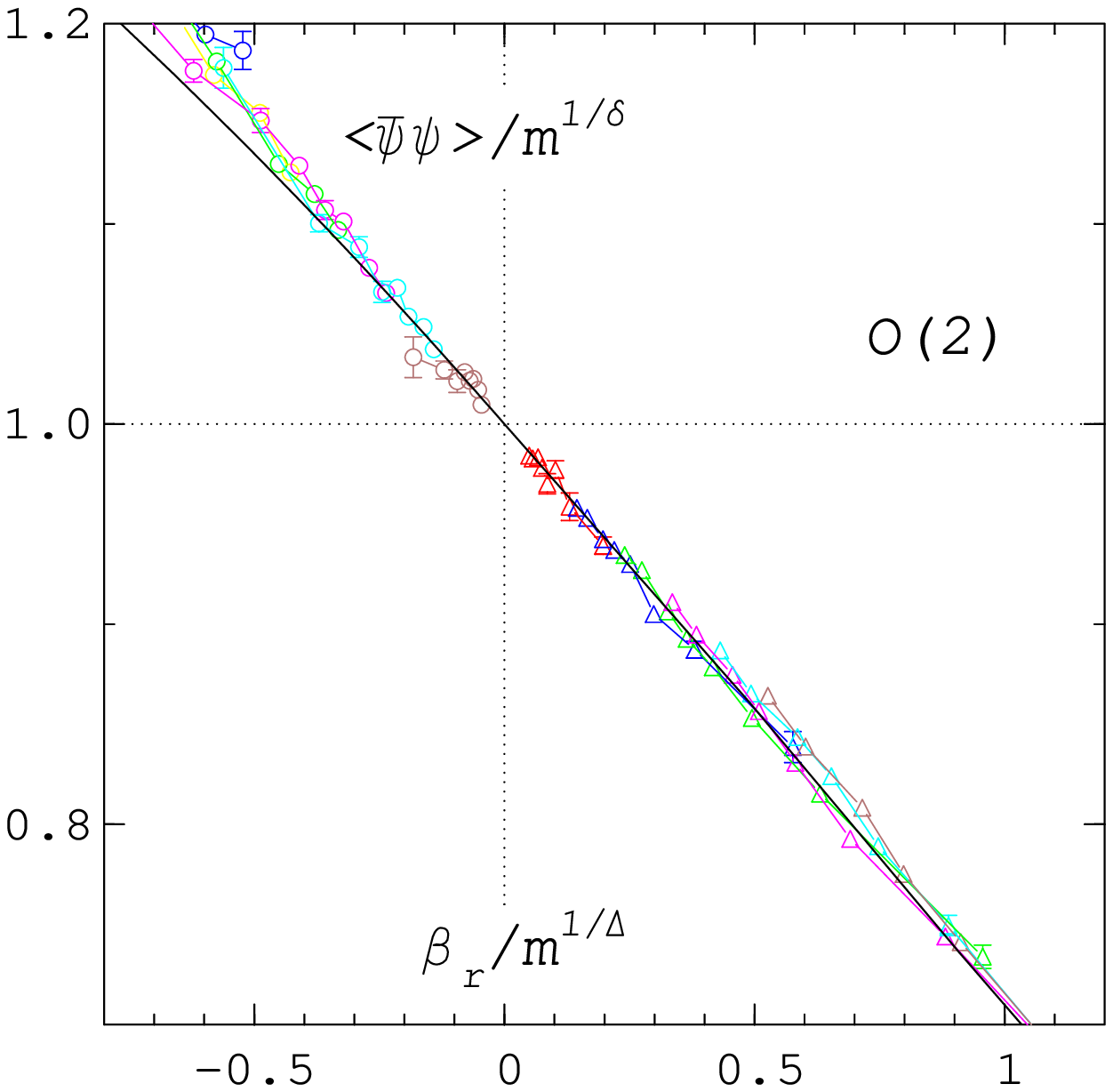, width=65mm}
          }
\put(8.1,0.5){
   \epsfig{bbllx=127,bblly=264,bburx=451,bbury=587,
       file=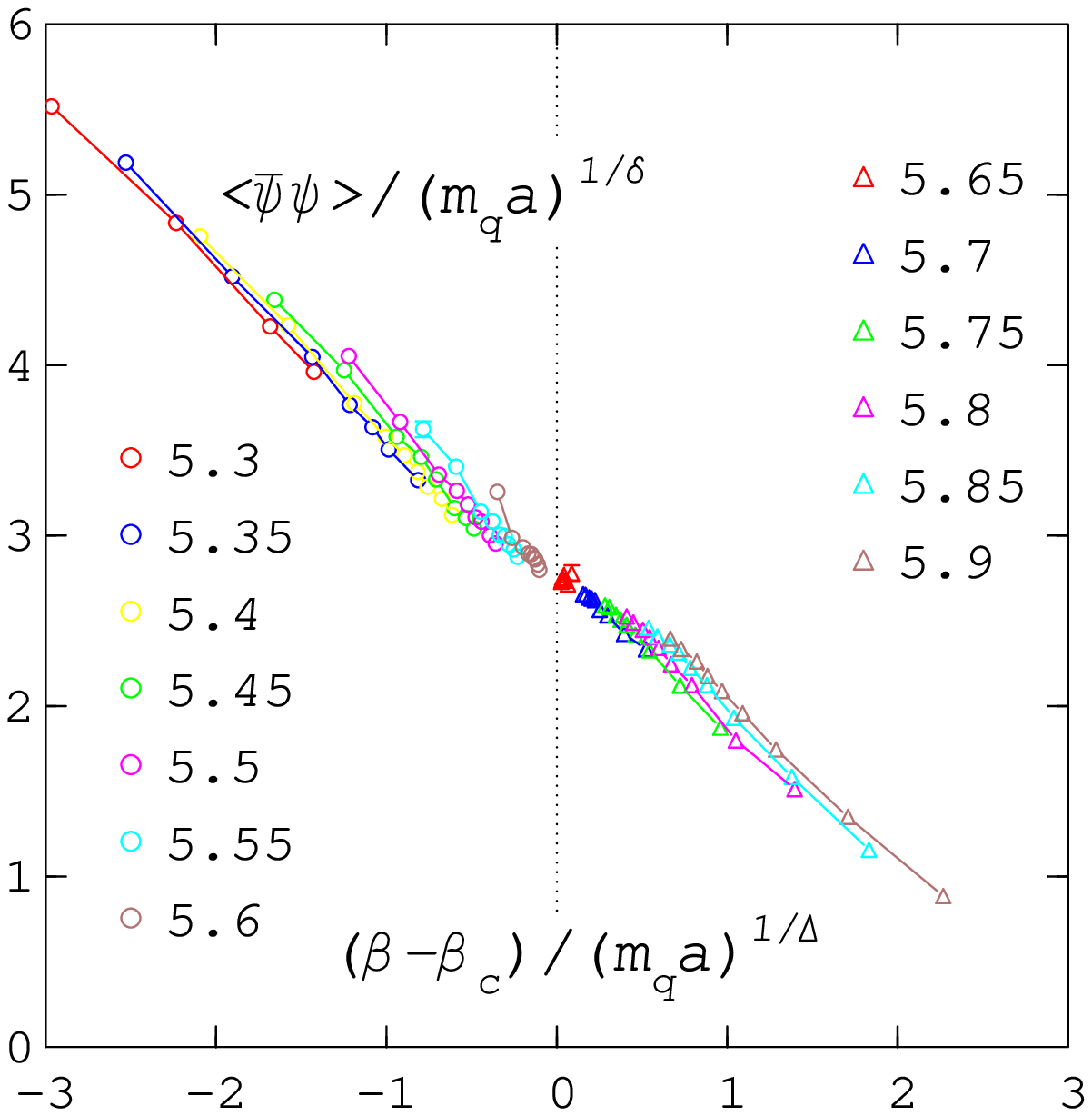, width=65mm}
          }
\end{picture}
\caption{Scaling test in $aQCD$. Left part: $\PBP/m^{1/\delta}$
   for $O(2)$ parameters in the vicinity of the critical point. The black
   line is the $O(2)$ scaling function. Right part: the unnormalized 
   variables with $\delta=4.4,\nu=1.0$ from Ref.\ \cite{Basile:2004wa}.
   The data for fixed $\beta$-values are connected by straight lines to
   guide the eye.}   
\label{fig:o2et}
\end{figure}

 We have performed explicit scaling tests for the data from the 
 $8^3\times 4$ lattice (they coincide with the results from the $\ns=12$
 and 16 lattices). In the left part of Fig.\ \ref{fig:o2et} the data for
 $\PBP m^{-1/\delta}$ are shown as a function of the scaling variable 
 $z=\beta_r /m^{1/\Delta}$ using $O(2)$ exponents.  The normalizations
 $(m_qa)_0$ and $\beta_0$ have been determined from the behaviour at
 $\beta_c$ (\ref{fitbc}), and the extrapolations $\PBP(\beta,0)$ obtained
 from the Goldstone effect \cite{Engels:2005te}. In addition we show the 
 $O(2)$ scaling function. We observe that the data scale very well 
 with $O(2)$ exponents in the small $|z|$-region and agree there with
 the $O(2)$ scaling function. For decreasing $z<0$ outside the shown range
 the data for different $\beta$ start to deviate and exhibit
 corrections-to-scaling as in the original $O(2)$ model. In the right part
 of Fig.\ \ref{fig:o2et} we show the best scaling result for the
 unnormalized variables obtained by varying the exponents around
 $\delta\approx 4.0,\nu\approx 1.1$, the expected continuum class values
 \cite{Basile:2004wa}. Here, $\beta_c=5.64$. We see from Fig.\ \ref{fig:o2et}
 that the data still spread considerably, a scaling function is not known 
 up to now. 
\newpage

\section{QCD with fundamental fermions}

Since the predictions for the universality class of ordinary $QCD$
could not be confirmed with the standard staggered action we use here
the $p4$-action\cite{Heller:1999xz}, which improves the cut-off 
dependence, rotational invarian\-ce and flavour symmetry. The
thermodynamics of two flavour $QCD$ has been investigated with this 
action in Ref.\ \cite{Karsch:2000kv}. In particular, the chiral
transition point was estimated to $\beta_c=3.48(3)$ by
extrapolation of the pseudocritical points $\beta_{pc}(m_qa)$ to $m_qa=0$.

We have extended the work of Karsch et al.\cite{Karsch:2000kv} on 
$\ns^3\times 4$ lattices with $\ns=8,12,16$ at the coup\-lings 
$\beta=2.8,3.0,3.2,,3.4,3.48$ and 3.50, that is for $\beta \le \beta_c$. 
All our simulations were done with the MILC code
using the R-algorithm with a mass-dependent
stepsize $\delta \tau(m_qa)=\min\{0.4m_qa,0.1\}$ and trajectory length
$\tau=1$. We produced 1000-2000 trajectories for masses $m_qa\in [0.025,0.5]$
and somewhat less for $m_qa=0.01$. 

\begin{figure}[t]
\begin{center}
      \epsfig{bbllx=63,bblly=265,bburx=516,bbury=588,
       file=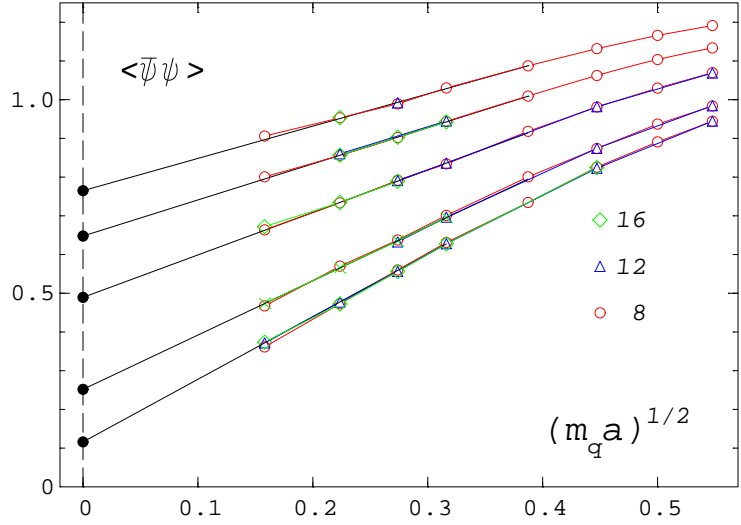,width=90mm}
\end{center}
\vspace*{0.2cm}
\caption{The chiral condensate $\PBP$ as a function of $(m_qa)^{1/2}$ for
all $\beta$-values between 2.8 (highest values) and 3.48 (lowest values)
from  $\ns^3 \times 4$ lattices with $\ns=8$ (circles), 12 (triangles) and
16 (diamonds). The black lines show the fit results and the filled circles 
the estimates for the chiral condensate at $m_qa=0$.}
\label{fig:geqcd}
\end{figure}

Like in $aQCD$ we have tested the mass dependence of the chiral condensate
at fixed $\beta<\beta_c$. In Fig.\ \ref{fig:geqcd} we show the data from 
$\beta= 2.8\,$ to $\beta=3.48\,$ as a function of $(m_qa)^{1/2}$. We have
fitted the data in the range $m_qa\lsim 0.15$ to the ansatz
\bq
\PBP(\beta,m_qa) = \PBP(\beta,0) +c_1(\beta) (m_qa)^{1/2}~.
\eq
Again, the behaviour expected due to the Goldstone effect is confirmed by
the data.

\begin{figure}[!ht]
\setlength{\unitlength}{1cm}
\begin{picture}(13,7)
\put(0.7,0.5){
   \epsfig{bbllx=127,bblly=264,bburx=451,bbury=587,
       file=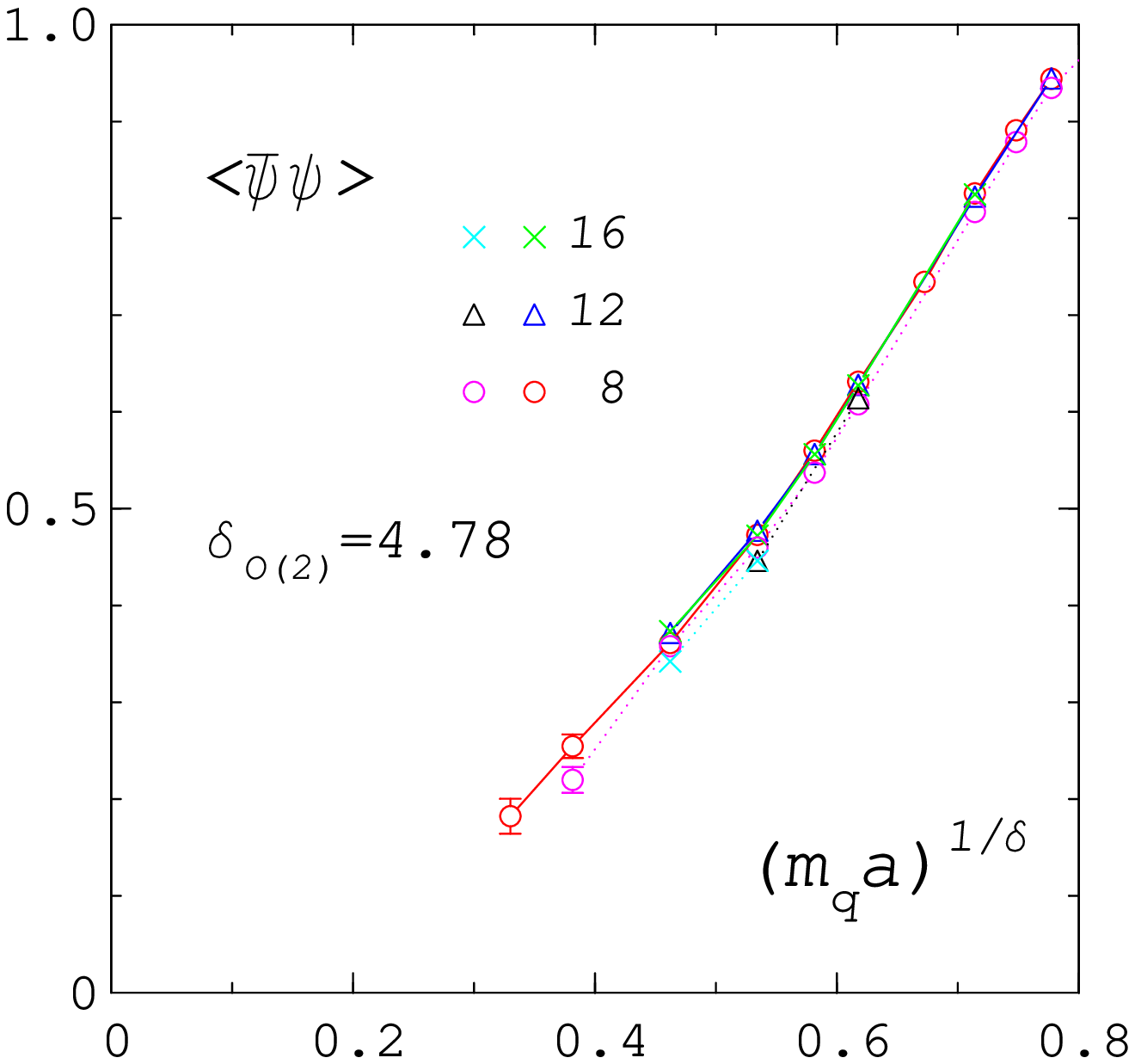, width=60mm}
          }
\put(8.3,0.5){
   \epsfig{bbllx=127,bblly=264,bburx=451,bbury=587,
       file=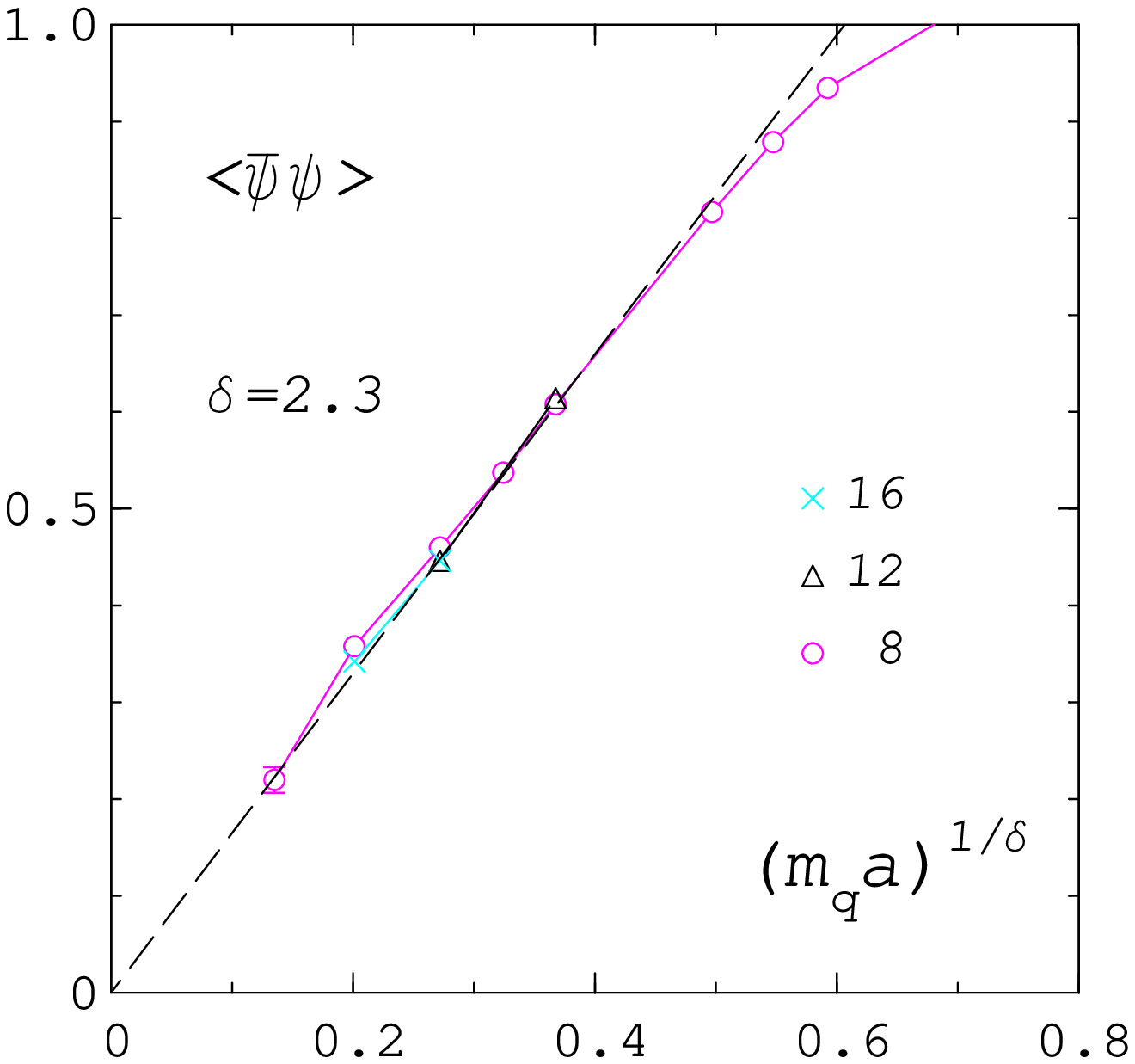, width=60mm}
          }
\end{picture}
\caption{Test of the scaling ansatz for $\PBP$ at the critical point 
   for $O(2)$ (left part) at $\beta=3.48$ (solid lines) and 3.5 (dotted 
   lines). In the right part we show a free fit to the scaling ansatz 
   at $\beta=3.5$ (dashed line). Here, the exponent is $\delta=2.3$.}   
\label{fig:pbpjc}
\end{figure}

\begin{figure}[t]
\setlength{\unitlength}{1cm}
\begin{picture}(13,7)
\put(0.6,0.5){
   \epsfig{bbllx=127,bblly=264,bburx=451,bbury=587,
       file=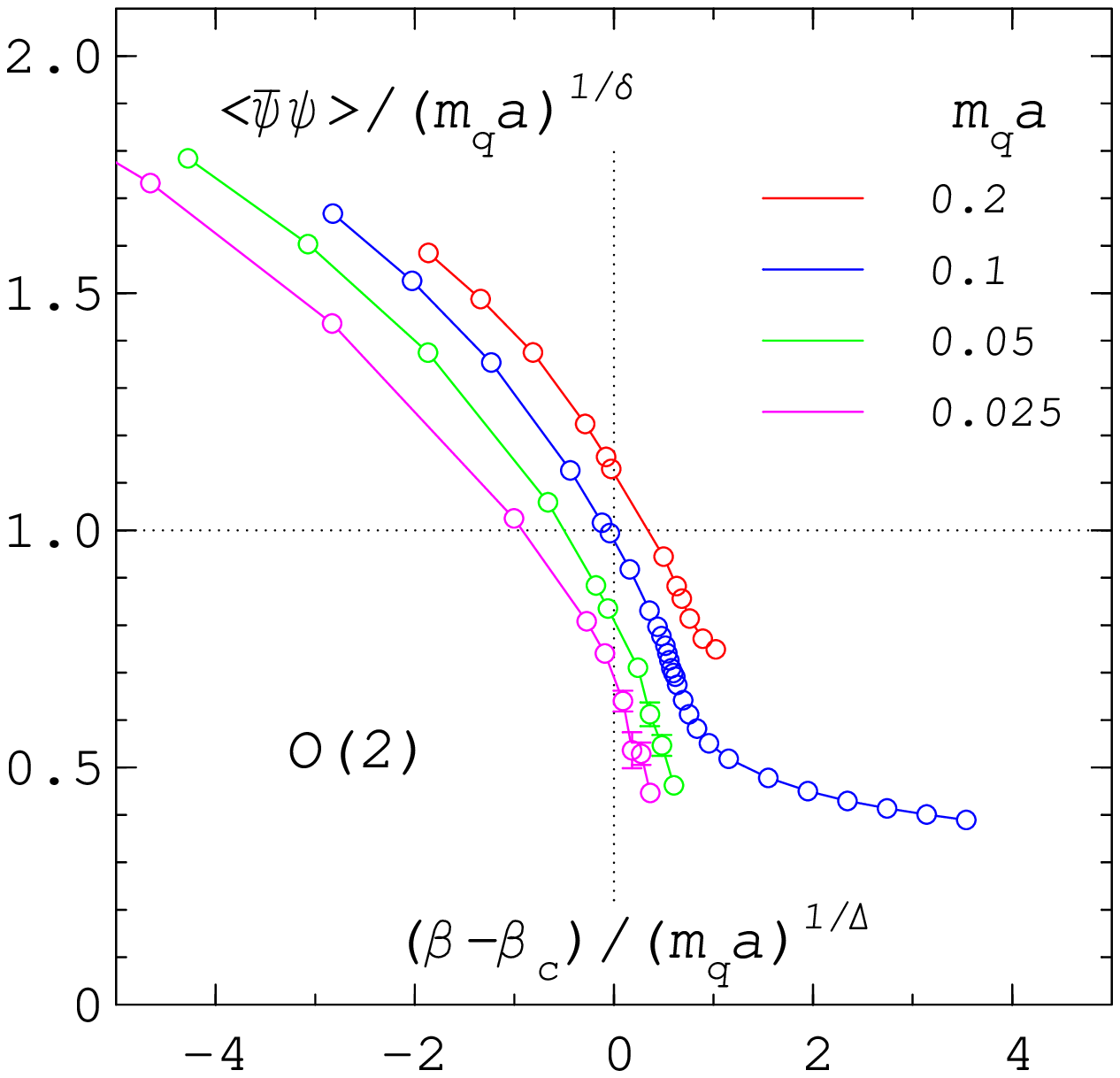, width=65mm}
          }
\put(8.2,0.5){
   \epsfig{bbllx=127,bblly=264,bburx=451,bbury=587,
       file=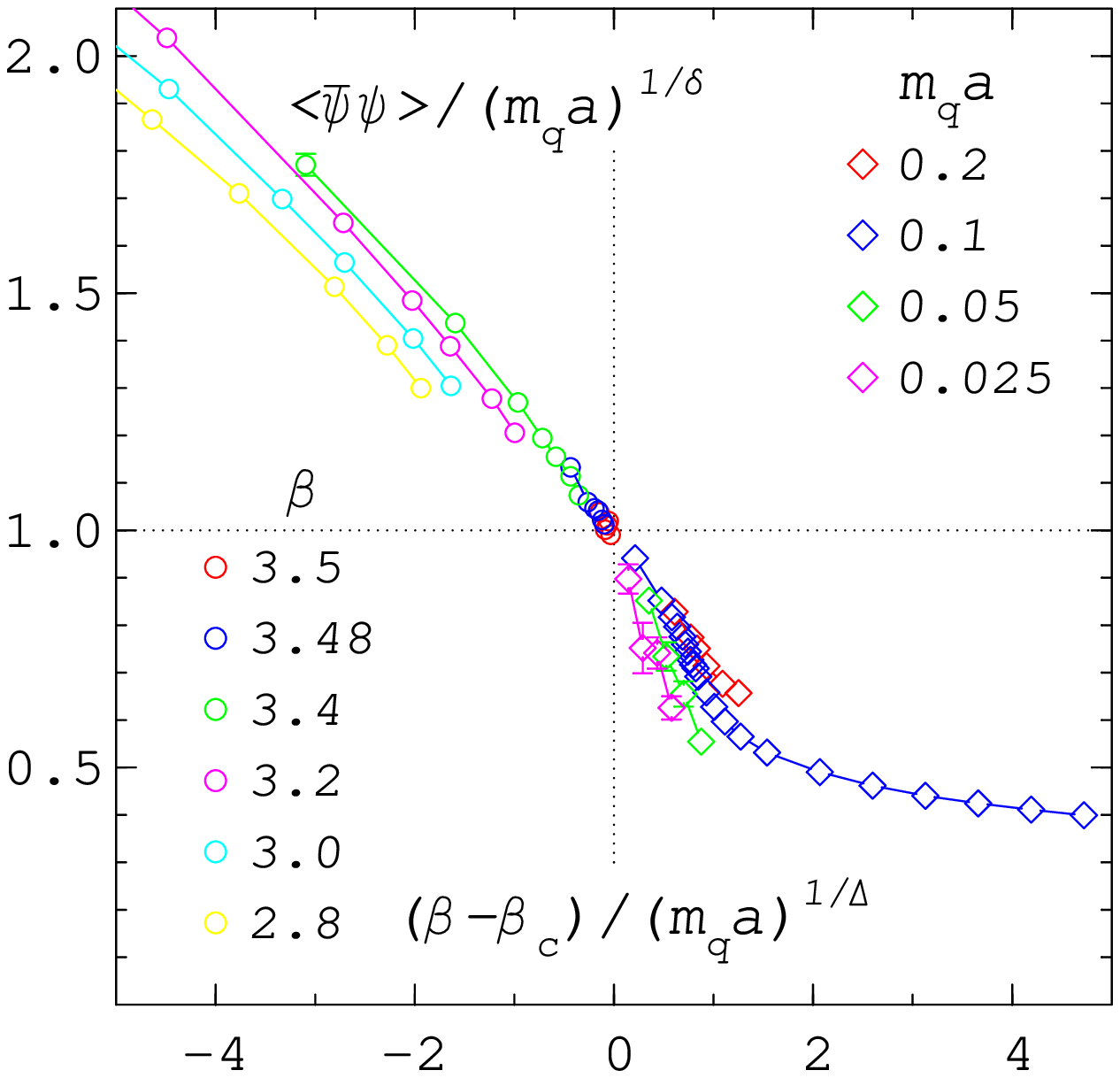, width=65mm}
          }
\end{picture}
\caption{Test on scaling behaviour of $\PBP/(m_qa)^{1/\delta}$ in $QCD$
   for $O(2)$ (left) and free (right) exponents in the critical region. 
   The data for fixed $\beta$ or fixed $m_qa$ are connected by straight
   lines to guide the eye.}   
\label{fig:scale}
\end{figure}

In Fig.\ \ref{fig:pbpjc} we investigate the scaling form $\PBP=d_c(m_qa)
^{1/\delta}$ for the critical point in the close neighbourhood of $\beta_c$.
We have plotted the data with $\delta_{O(2)}=4.78$ in the left part of 
Fig.\ \ref{fig:pbpjc} at $\beta=3.48$ (solid lines) and $\beta=3.5$
(dotted lines). Obviously, the curves are no straight lines going through
the origin. Also, the corresponding attempt to obtain a scaling function
with $O(2)$ exponents fails as can be seen in the left part of Fig.\ 
\ref{fig:scale}. Here, $\beta_c=3.51$ was used, the scales are not 
normalized. Other values for $\beta_c$ shift the results slightly, but do
not lead to scaling. The use of $O(4)$ exponents does not improve the result. 

In order to achieve scaling of the data we have fitted the ansatz 
$\PBP=d_c(m_qa)^{1/\delta}$ at $\beta=3.50$ with a free amplitude $d_c$ 
and a free exponent $\delta$ for $m_qa\in [0.01,0.1]$ and obtain
$\delta=2.3$. The fit is shown in the right part of Fig.\ \ref{fig:pbpjc}.
Likewise, we estimated the magnetic exponent $\beta_m$ to 0.6 from the
values of $\PBP$ at $m_qa=0$ which we gained from the Goldstone extrapolations. 
In the right part of Fig.\ \ref{fig:scale} we show the data for
these exponents. Here, $\PBP/(m_qa)^{1/\delta}$ is normalized to 1 at 
$\beta_c$. Still, scaling is not perfect, but certainly much better than
for $O(N)$ exponents.
\newpage
\section{Summary}

In $aQCD$ the deconfinement transition is first order and occurs below
the second order chiral transition. The latter is located at
$\beta_c=5.624(2)$. The scaling behaviour of $\PBP$ in the vicinity 
of $\beta_c$ is in full agreement with the $3d$ $O(2)$ universality class.
The lattice data do not yet show scaling with exponents from the proposed
continuum class\cite{Basile:2004wa}. In the 
region between the two phase transitions the quark mass dependence of
the chiral condensate is as expected due to the exis\-tence of Goldstone
modes like in $3d$ $O(N)$ spin models.

For $QCD$ we find that the chiral condensate exhibits Goldstone effects 
below the chiral transition point as in $3d$ $O(N)$ spin models and like
in $aQCD$. The transition is of second order or a crossover, there is no
sign of a first order behaviour. In the vicinity of $\beta_c$ the chiral
condensate does not show the scaling behaviour of the $3d$ $O(2)$ or $O(4)$
universality classes, at best it is that of a class with exponents 
$\delta=2.3$ and $\beta_m=0.6$. This result could be a consequence
of the coincidence of the deconfinement and the chiral transitions.

\end{document}